\documentclass[aps,rmp,12pt,preprint]{revtex4}
\usepackage{natbib}
\usepackage{amssymb}
\usepackage{graphicx,epsfig,chemarr}
\usepackage[usenames,dvipsnames]{color}
\usepackage{times}
\usepackage{verbatim}
\usepackage{amsmath}
\usepackage{graphicx}
\usepackage{amssymb}
\usepackage{epsfig}

\def\JPE#1{JPE: \color{red} {\tt #1}}

\def\be{\begin{equation}}
\def\ee{\end{equation}}
\def\ba{\begin{eqnarray}}
\def\ea{\end{eqnarray}}
\def\kbar{\bar{k}}

\def\JPE#1{\kern 30ptJPE:\ \color{red}{\tt #1}\color{black}}

\def\ie{{\it{i.e.}}}

\def\fs{{f_\ast}}

\def\ts{{t_\ast}}

\def\erf {{\mathrm{erf}}}
\def\const{{\mathrm{const.}\,}}
\def\kmin {k_{\rm{min}}}
\def\kmax {k_{\rm{max}}}
\def\ktail {k_{\rm{tail}}}
\def\kin{k_{\rm{in}}}
\def\kout{k_{\rm{out}}}
\def\Phitail {\Phi_{\rm{tail}}}
\def\ltail{\ell_{\rm{tail}}}
\def\lmax {\ell_{\rm{max}}}

\begin{document}

\title{Leaders of neuronal cultures in a quorum percolation model}

\author{J.-P. Eckmann$^{1,2}$}
\author{Elisha Moses$^{3,*}$}
\author{Olav Stetter$^4$}
\author{Tsvi Tlusty$^3$}
\author{Cyrille Zbinden$^1$}

\affiliation{$^1$ D\'{e}partement de Physique Th\'{e}orique and
$^2$ Section de Math\'{e}matiques, Universit\'{e} de
Gen\`{e}ve, CH-1211 Gen\`{e}ve 4, Switzerland,}
\affiliation{$^3$ Department of Physics of Complex Systems, Weizmann
Institute of Science,  Rehovot, Israel} \email{
elisha.moses@weizmann.ac.il} \affiliation{$^4$ Max Planck Institute
for Dynamics and Self-Organization, and Bernstein Center for
Computational Neuroscience, Bunsenstr.  10, 37073 G\"{o}ttingen,
Germany}

\begin{abstract}
We present a theoretical framework using quorum-percolation for
describing the initiation of activity in a neural culture. The
cultures are modeled as random graphs, whose nodes are excitatory
neurons with $\kin$ inputs and $\kout$ outputs, and whose input
degrees $\kin=k$ obey given distribution functions $p_k$. We examine
the firing activity of the population of neurons according to their
input degree ($k$) classes and calculate for each class its firing
probability  $\Phi_k(t)$ as a function of $t$. The probability of a
node to fire is found to be determined by its in-degree $k$, and the
first-to-fire neurons are those that have a high $k$. A small
minority of high-$k$ classes may be called ``Leaders,'' as they form
an inter-connected subnetwork that consistently fires  much before
the rest of the culture. Once initiated, the activity spreads from
the Leaders to the less connected majority of the culture. We then
use the distribution of in-degree of the Leaders to study the growth
rate of the number of neurons active in a burst, which was
experimentally measured to be initially exponential. We find that
this kind of growth rate is best described by a population that has
an in-degree distribution that is a Gaussian centered around $k=75$
with width $\sigma=31$ for the majority of the neurons, but also has
a power law tail with exponent $-2$ for ten percent of the
population. Neurons in the tail may have as many as $k=4,700$
inputs. We explore and discuss the correspondence between the degree
distribution and a dynamic neuronal  threshold, showing that from
the functional point of view, structure and elementary dynamics are
interchangeable. We discuss possible geometric origins of this
distribution, and comment on the importance of size, or of having a
large number of neurons, in the culture. \vspace{0.5cm}

\end{abstract}

\maketitle

\section{Introduction}
Development of connectivity in a neuronal network is strongly
dependent upon the environment in which the network grows: cultures
grown in a dish will develop very differently from networks formed
in the brain. In a dish, the only signals that neurons are exposed
to are chemicals secreted by neighboring neurons, which then must
diffuse to other neurons via the large volume of fluid that
surrounds the culture. The result is a connectivity dominated by
proximity in a planar geometry, whose input degree follows a
statistical distribution function that is Gaussian-like \cite{PNAS}.
This is contrasted by the intricate guidance of axons during the
creation of connectivity in the brain, which is dictated by a
detailed and very complex ``blueprint'' for connectivity. As a
result, the firing pattern of a culture is an all-or-none event, a
population spike in which practically all the neurons participate
and are simultaneously active for a brief period of time, spiking
about 3-4 times on average.

We have previously shown \cite{EPL} that graph theory and
statistical mechanics are useful in unraveling properties of the
network in a rat hippocampal culture, mostly because of the
statistical nature of the connectivity. With these tools we have
been able to understand such phenomena as the degree distribution of
input connections, the ratio of inhibitory to excitatory neurons and
the input cluster size distribution \cite{PRL,PNAS}. We found that a
Gaussian degree distribution gives a good quantitative description
for statistical properties of the network such as the appearance of
the giant $m$-connected component and its size as a function of
connectivity. The inhibitory component was found to be about $30\%$
in hippocampal cultures, and about $20\%$ in cortex. We furthermore
observed that the appearance of a fully connected network coincides
precisely with the time of birth \cite{PNAS}.

In this paper we apply our graph theoretic approach to the
intriguing process of the initiation of a spontaneous population
spike. On the one hand, a perturbation needs to be created that
pushes a number of neurons to begin firing. On the other hand, the
initial firing pattern must propagate to the rest of the neurons in
the culture. Understanding this recruitment process will give
insight on the structure of the network, on the interrelation of
activation in neurons, and on the dynamics of neuronal firing in
such a complex culture. A simple scenario for initiation that one
might conceive of is wave-front propagation, in which a localized
and limited area of initiation is ignited first, and from there sets
up a spherical traveling front of excitation. However, as we shall
see, the initiation is a more intricate process.

The experimental situation regarding initiation of activity is
complex. In quasi one-dimensional networks we have been able to show
that activity originates in a single ``Burst Initiation Zone''
(BIZ), in which a slow recruitment process occurs over several
hundreds of milliseconds
\cite{OferJNP1,OsanErmentrout,GolombErmentroutPNAS,GolombErmentroutPRE}.
From this BIZ the activity propagates to the rest of the linear
culture along an orderly and causal path dictated by the one
dimensional structure\cite{OferJNP2}. In two dimensional networks
such causal propagation is not observed \cite{MaedaJN,StreitEJN},
and the precise mode of propagation has not been identified.
Recently, Eytan and Marom \cite{EandMJN} found that a small subgroup
of neurons were the ``first-to-fire,'' and that also in this case
the initiation is long (on the order of hundreds of milliseconds). They also
observed that the growth rate is exponential in the initial stage and
then changes to faster than exponential. These neurons were later
shown to characterize and ``lead'' the burst \cite{NJP}, and to
recruit the neurons in their proximity in the ``pre-burst'' period \cite{NJP}.

In this paper we address and connect three experimental observations
that are at first sight unrelated. The first and fundamental
observation is the fact that bursts are initiated by Leaders, or
first-to-fire neurons \cite{EandMJN}. We use the quorum-percolation
model to answer the question -- what makes these Leaders different
from the rest of the network -- by showing that one of their
important characteristics is a high in-degree, \ie, a large number
of input connections.

We then turn to the second experimental observation, that the
activity in a burst starts with an exponential growth. We show that
this can happen only if the distribution of in-degree is a power law
with exponent of $-2$. However, this needs to be related with the
third experimental observation, which is that the distribution of
in-degrees is Gaussian rather than power law. We reconcile both
observations by stitching together the two solutions into an
in-degree distribution that has the large majority of neurons in a
Gaussian centered around an average in-degree of about $75$, while
ten percent of the population lie on a power law tail that can reach
a few thousand connections. We show that this reconstructs the full
experimentally observed burst structure, which is an exponential
initiation during the pre-burst followed by a super-exponential
during the burst.

We  present several ideas on the origin of these
distributions in the spatial extent and geometry of the neurons, and
show that the distribution of in-degrees is proportional to the
distribution of spatial sizes of the dendritic trees. We thus
conjecture that the distribution of dendritic trees is mostly
Gaussian, but that a few neurons must have dendrites that go off
very far, with power law distribution of this tail. We end by making
some additional conjectures about one-dimensional cultures and on
the importance of the size of the culture.

\section{Methods}
\subsection{Quorum percolation model for dynamics of random graph network}

We describe the neural culture using a simplified model of a network
whose nodes are neurons with links that are the axon/dendrite
connections. This picture is further simplified if we consider a
randomly connected sparse graph \cite{Bollobas}, with a uniform
strength on all the connections. The structure and topology of the
graph are determined by specifying a probability distribution $p_k$
for a node to have $k$ inputs. Percolation on a graph is the process
by which a property spreads through a sizeable fraction of the
graph. In our case, this property is the firing of neurons. The
additional characteristic of Quorum Percolation is, as its name
implies, that a burst of firing activity will propagate throughout
the neuronal culture only if a quorum of more than $m$ firing
neighbors has ignited on the corresponding graph. While this
description makes a number of assumptions that are not exact in
their comparison to the experiment, it does, as we have been able to
show previously, capture the essential behavior of the network
\cite{EPL,Remark}. The use of such a simple model for the neuronal
network is justified at the end of the Methods section.

In particular, in this paper we obtain a theoretical explanation for
the experimental observation of initiation of activity by a small
number of neurons from the network and the subsequent gradual
recruitment of the rest of the network. Within the framework of a
random graph description, we have previously shown that the dynamics
of firing in the network is described by a fixed point equation for
the probability of firing in the network, which also corresponds to
the experimentally observed fraction of neurons that fire.
Experimentally, this fraction can be set by applying an external
voltage \cite{PRL,PNAS}. In the case of spontaneous activity, this
fraction is determined by the interplay of noise and the intrinsic
sensitivity of the neurons \cite{EnricJCN}.

Specifically, connections are described by the adjacency matrix $A$,
produced according to the probability distribution $p_k$, with
$A_{ij}=1$ if there is a directed link from $j$ to $i$, and
$A_{ij}=0$ otherwise and $A_{ii}=0$.

Our study starts by assuming initial conditions where a fraction $f$
of neurons are switched ``on'' externally at time $t=0$. The neurons
fire, and once they do they stay "on" forever - a neuron will be on
at time $t+1$ if at time $t$ it was on. A neuron is either turned on
at $t=0$ (with probability $f$) or, if it is off at time $t$, then
it will it will be turned on at time $t+1$ if at least $m$ of its
upstream (incoming) nodes were on at time $t$: \be
s_{i}(t+1)=s_{i}(t)+ \left( 1-s_{i}(t)\right) \Theta \left(
\sum_{j}A_{ij}s_{j}(t)-m\right) , \label{dynamics} \ee where
$s_{i}(t)$ describes the state of the neuron at time $t$ (1 for
``on'' and 0 for ``off'') and $\Theta$ is the step function (1 for
$x \ge 0$, 0 otherwise). Note that the second term, which accounts
for the neuron's probability to fire at time step $t$, creates a
coupling of $s_i(t)$ to all its inputs $s_j(t)$. Lacking a turning
off process, the number of active nodes is  monotonically increasing
and converges  into a steady-state (a fixed point) within a finite
time $t_{f}$, which is smaller than the number of neurons,
$t_{f}<N$.

To derive the ``mean-field'' dynamics for the average fraction of
firing neurons, $\Phi(t)=\left\langle s_i(t) \right\rangle =
 N^{-1}\sum_i s_i(t),$ one cannot simply average
equation  (\ref{dynamics}) directly. This is due to the correlation
between $s_i(t)$ and $\Theta\left(\sum_{j}A_{ij}s_{j}(t)- m
\right)$. The correlation exists because if a given neuron fires,
$s_i(t)=1$, and it was not externally excited, then at least $m$ of
its inputs are firing and the step function over its inputs must
also be $1$.  In fact, at the  fixed point the correlation is
strong, $\left(
1-s_{i}(t)\right)\Theta\left(\sum_{j}A_{ij}s_{j}(t)-m\right) = 0$,
since in the steady state a neuron can remain off if and only if
less than $m$ of its inputs fire. To avoid the correlations, one
utilizes the monotonicity to realize that a neuron is on only if it
was turned on externally at time $t=0$ or, if it was off at $t=0$
then at some time $t$ at least $m$ of its inputs fired. We can then
replace $s_i(t)$ by $s_i(0)$ and rewrite equation (\ref{dynamics})
as $ s_{i}(t+1)=s_{i}(0)+ \left( 1-s_{i}(0)\right) \Theta\left(
\sum_{j}A_{ij}s_{j}(t)-m\right)$. In the tree approximation, which
disregards loop feedbacks, the initial firing state of a neuron,
$s_i(0)$, cannot affect its inputs, $s_{j}(t)$. Inversely, it is
obvious that $s_i(0)$ which is determined externally is independent
of $s_j(t)$. Therefore, $s_i(0)$ and $\Theta\left(
\sum_{j}A_{ij}s_{j}(t)-m\right)$ can be averaged independently. The
result is the mean-field iteration map
\be \Phi(t+1) =
f+(1-f)\Psi(m,\Phi(t))~, \label{DynamicMaria} \ee
where the
combinatorial expression for $\Psi $ is the probability that at
least $m$ inputs are firing and $f$ is the initial firing fraction,
$f = \Phi(0)=\left\langle s_i(0) \right\rangle$.
 The steady state of the network is defined by the fixed point
$\Phi_\infty$, which is found by inserting
$\Phi(t+1)=\Phi(t)=\Phi_\infty$ into equation (\ref{DynamicMaria}),
to obtain $\Phi_\infty = f+(1-f)\Psi(m,\Phi_\infty)$.

\subsection{Collectivity and the critical point}

The role of Leader neurons in the initiation and the development of
bursts can be clarified by dividing the neurons into classes of
in-degree $k$ (``$k$-class'') and looking at the dynamics of each
class separately. The total firing probability $\Phi = \sum_k p_k
\Phi_k$ is thus composed of the sum over the individual
probabilities $\Phi_k$ for each $k$-class to fire. The mean field
equation for a given $k$-class is \be \Phi_k(t+1) =
f+(1-f)\Psi_k(m,\Phi(t))~, \label{kDynamicMaria} \ee where $\Psi_k $
is the probability that a neuron with $k$ inputs has at least $m$
that are firing. Although all $k$-classes are coupled through the
common $\Phi$, the formulation of equation \ref{kDynamicMaria}
allows the tracing of the fraction $\Phi_k$ of each class and its
dynamics during its evolution.

It follows
from (\ref{kDynamicMaria}) that at the fixed point $\Phi_{k,\infty}
= f+(1-f)\Psi_k(m,\Phi_\infty)$. Given the time dependence of $\Phi(t)$, one
can extract the fraction of firing neurons in each $k$-class, $\Psi_k$, by
plugging $\Phi(t)$ into
(\ref{kDynamicMaria}).

The combinatorial expressions for
$\Psi$ and $\Psi_k$ are:
\be \Psi_k(m,\Phi) = \sum_{l=m}^{k } \dbinom{k}{l} \,\Phi^l \,
(1-\Phi)^{k-l}\; ;\;\;\;\;\;
 \Psi(m,\Phi)=\sum_{k=0}^{\infty} p_k \Psi_k(m,\Phi)~.
 \label{cpmb}\ee


There is a particular critical initial firing $\fs$ where the
solution jumps from $\Phi \approx f$ (\ie, practically all
activation is externally driven and there is almost no collectivity,
$ \Psi \ll 1$) to $\Phi \approx 1$ where most firing is due to
inputs (and $\Psi \simeq 1$). It is both convenient and instructive
to treat and simulate the network near this transition point, since
the dynamics there is slow. This allows the different steps in the
recruitment process to be easily identified and distinguished.

\subsection{Simulation of the quorum percolation model}

The model was numerically investigated by performing by a
simulation, employing $N=500,000$ neurons. This number was chosen to
match as close as possible the number of neurons typically in an
experiment, which is on the order of one million. An initial
fraction $f$ of the neurons were randomly selected and set to ``on''
(\ie, fired). At every time step a neuron would fire if it fired the
previous step, or if more than a threshold number of its input
neighbors fired. The threshold $m$ and the initial firing component
$f$ could be varied, and the activity history of all neurons was
stored for subsequent analysis. We used a number of different degree
distributions, including a Gaussian, exponential and power law for
the network. We also used a tailored Gaussian distribution in which
$10\%$ of the high $k$ neurons, which have $k$ higher than a given
$k_{\rm{tail}}$, obey a power law distribution function.

\subsection{Validity of the model}

\subsubsection{The use of a random graph for neuronal cultures}
The spatial extent and arrangement of connections can be of
importance to the dynamics of the network. In contrast to spatially embedded
(metric) graphs, random graphs allow any two nodes to connect, \ie they are
the analog of infinite dimensional networks.  The experiment is obviously
metric but our model employs a random graph. This seeming contradiction was
resolved  in a previous study \cite{Remark}, where we showed that if
the average connectivity is high enough then the graph is
effectively random (i.e. of very high dimension).
Why does the random graph picture describe so successfully the measurements of a 2D neural culture while it completely neglects the
notions of space and vicinity?  As we explained in \cite{Remark}, there is a basic difference in the
manner in which random and metric graphs are ignited. In metric graphs it suffices to initially turn
on localized excitation nuclei, which are then able to spread an excitation front throughout
the spatially extended network. In random graphs, there are no such nuclei  and one has to excite a finite fraction
of the neurons to keep the ignition going. Still, we showed \cite{Remark}
that  the experimental network,
which is obviously an example of a metric graph, is effectively random, since its finite size
and the demand for a large quorum of firing inputs makes the occurrence of excitation nuclei very improbable. As explained in
that paper, this occurs when it becomes impossible to identify
causal paths in space along which activity propagates, with one
neuron activating its neighbor and so on. In other words, the
activity burst does not initiate at one specific nucleating site,
and has instead multiple locations at which activity appears. This
is exactly the characteristic of a random graph with no spatial
correlations. Thus a highly connected graph in 2D such as ours has
characteristics that are similar to a high dimensional graph with
near-neighbor connectivity.

\subsubsection{Approximating neuronal cultures as tree-like graphs}
The basic reason why a tree-like graph will describe the
experimental network lies in the observation that the percentage of
connections emanating from a neuron that actually participate in a
loop is small. Indeed, we have demonstrated in \cite{PNAS} that the
average number of connections per neuron is large, about $100$. On
the other hand, because the network is built from dissociated
neurons, the connectivity is determined by a spatial search process
during their growth, which is for all practical purposes a random
one. We therefore have a random network (embedded in a metric space)
with about $100$ connections per node.

Such networks do indeed have some loops, and thus we need to study
the effect of such loops on the general argument of equation
\ref{DynamicMaria}. For this, it suffices to study the case of
$2$-loops (which in fact cause the strongest correlations). Assume
neuron A and neuron B are linked in a $2$-loop.

If neuron A fires at time $t$ then it does not change any more, and
therefore the state of B at time $t+1$ does not matter for the state
of A at any later time. If A is not firing at time $t$ then it
decreases the probability of B to fire at time $t+1$, and this in
turn reduces its own probability to fire at time $t+2$. Clearly,
this effectively decreases the probability of A to fire. We shall
now show that this effect is $1/(k^2N)$ where $k$ is the mean degree
($100$) and $N$ is the total number of neurons that B can connect
to.

To see this, we note that if A is off then the number of available
inputs that can fire into B reduces by one, from $k$ to $k-1$. We
show below that this corresponds changing the ignition level $\Phi$
from $m/k$ to $m/(k-1)$. The overall effect of a $2$-loop on the
probability of B to fire therefore scales like $m/(k-1)-m/k\simeq
m/k^2$. The back-propagated effect on A will be of order $m^2/k^4$.

To estimate the total number of $2$-loops that include neuron A we
first look at all trajectories of length $2$ that emanate from A.
There are $\kout^2$ such trajectories. Of these a fraction of
$\kin/N$ will return to A. The total number of $2$-loops that start
at A is thus $\kout^2\kin/N$. The fraction of inputs of A that
participate in a $2$-loop is therefore $\kout^2/N$

Assuming that for a typical neuron $\kin$ is equal to $\kout$, the
total effect of $2$-loops on $\Phi$ calculated at neuron A is
therefore $m^2k^2/N/k^4=m^2/(k^2N)$.

Below we show the applicability of the tree like random graph model
by comparing its results directly to the simulation that uses
$N=500,000$ neurons. The correspondence between model and
simulations (Figures \ref{f:ex4a} and \ref{maromia}) is satisfactory.

In the experimental case, spatial proximity may lead to more
connections than in a random graph. The effect of space is to change
the relevant number of neurons $N$ from the total number to those
that are actually accessible in 2D. That number $N$ is on the order
of $N_{\rm space} = 3,000$ as compared to $N_{\rm total}=500,000$.
However, $1/k^2N$ is still a small number.

In a separate work, a simulation that takes space into account was
performed \cite{ZbindenThesis}, and the number of loops could be
evaluated directly. Indeed we found that the number of loops is
enhanced over the random graph estimation by a factor of $N_{\rm
space}/N_{\rm total}$, but still remains small.

\subsubsection{Applicability of the averaged equation (mean-field) approach}
In a physical model one must be sure that the ensemble of random
examples chosen to average over a given quantity does indeed
represent well the statistics of the system that is being treated.
In our model, the connectivity of the graph is fixed (``quenched''
disorder), and the ensemble is that of the random graphs that can be
generated with the particular choice of input connection
distribution function. In reality, the experiment and the simulation
measure the bursts inside one particular realization. However, the
mean-field equation averages over a whole ensemble of such random
graphs. The question is whether the averages obtained using one real
graph are representative of the whole ensemble. This is a behavior
which is termed ``self-averaging'', and means that, in the limit of
large graphs, one single configuration represents the average
behavior of the ensemble.

The similar ensemble of the classic Hopfield
(spin-glass) model for neural networks is known to be self averaging
in the limit of an infinite sized graph
\cite{SompolinskyPRA1985,vanHemmenPRL1982,ProvostValleePRL1983}. This occurs because the
dynamics is performed over a huge number of single neuron
excitations.

In practice, our model differs from the neural network model of
\cite{SompolinskyPRA1985} in that the neurons of our model change
only from ``off'' to ``on'', and cannot flip in both directions as
the equivalent ``spins'' do. We therefore tested numerically the
self-averaging property of the graph, and found that it indeed
exists for the random graphs we considered \ie, the fluctuations
between specific different realizations of the graph are negligible (see
 Fig. \ref{f:ex4a} below).

\subsubsection{The model describes initial growth of activity}

The possibility of turning a neuron ``off'' is not incorporated in
the model because we only consider short times. The whole process
described by the simulation occurs over a very brief period of time,
and therefore a firing neuron keeps its effect on other neurons
during the whole process. To be concrete, the unit of time in the
model and in the numerical simulations is the firing of one spike,
equivalent in the experiment to about one millisecond. The
simulation extends to about $50$ units, i.e. describes a process
that occurs typically for $50$ ms.

In our model a neuron has no internal structure, so that whether it
is ``on'' or ``off'' impacts only the neurons that are its
neighbors. The relevant issue is therefore - how long is the effect
of a neuron's activity felt by its 'typical' neighbor. The
experimental facts are that a neuron fires on average 4-5 spikes per
burst, each lasting a millisecond, with about 4-5 milliseconds
between the spikes (\cite{ShimshonJNP}) so that its total active
time spans typically 20 milliseconds.

The post-synaptic neuron retains the input from these spikes over a
time scale set by the membrane decay constant, which is on the order
of 20-30 milliseconds. Therefore, after a firing period of about 20
milliseconds, there is a retention period of comparable duration. We
can conclude that the effect of a neuron that has fired can be felt
by its neighbors for the total build-up time of the burst, about 50
milliseconds. We therefore describe by ``on'' the long term,
averaged effect of the neuron once it has begun firing. One caveat
to this is that the strength of that effect may vary with time, and
such an effect is not described within the model.

We also assume, for simplicity, that all the neurons are available
and can participate in the burst (no refractive neurons). In the
experiment this is equivalent to looking at those bursts that have
quiescent periods before them, which is often the case.

\subsubsection{The role of inhibitory neurons}
In this model all neurons are excitatory; within the ``0'' or ``1''
structure of the model, the contribution of an inhibitory neuron
would be ``-1''. Thus adding inhibitory neurons amounts to
increasing $m$, the number of inputs that must fire before a neuron
will fire. This is a small accommodation of the model, and does not
change the dynamics of burst initiation.

Below we also model the dynamics of burst activation observed in the
experiments of Eytan and Marom \cite{EandMJN}, which measured an
exponential recruitment at the initial stage of the burst. These
experiments clearly show that the dynamics are essentially the same
for cultures both with and without inhibition. Both display an
initial exponential growth followed by a super-exponent. One small
difference lies in the value of the growth, which is larger for
dis-inhibited than for untreated cultures. But the main difference
is seen only after the burst reaches its peak, in the decay of the
burst.

\subsubsection{The role of noise in burst initiation}
We assume the existence of spontaneous sporadic activity of single
neurons in the culture. In principle, this can be treated as a
background noise \cite{EnricJCN}. In our case we require that a
minimal amount $f$ of the culture spontaneously fires, and we look
at the ability of this fraction of initially firing neurons to
initiate a burst. It is possible to initiate the activity with an
external voltage $V$, using bath electrodes, as we reported in
previous work (\cite{PRL, PNAS}). In that case $f(V)$ is determined
by the percentage of neurons that are sensitive enough to fire at a
voltage $V$.

\section{Results}
\subsection{First to fire neurons lead bursts and have large input
degree} \label{IIIA}

\begin{figure}[ht]
  \begin{center}
   \includegraphics[scale=0.6]{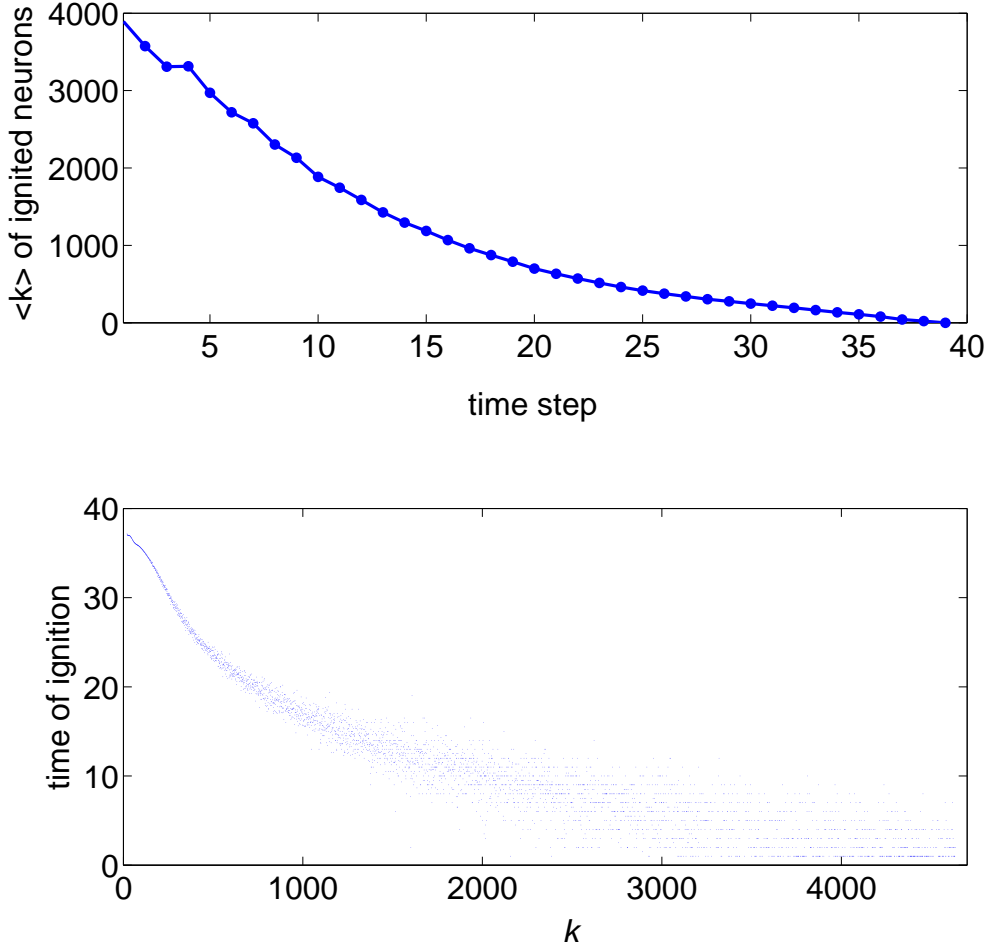}
  \caption{ Top: Average $k$ of all neurons ignited at each time step $t$
  from one particular realization of the simulation.
   Bottom: Times of ignition for all $500,000$ individual neurons.
   It is evident that highly connected neurons are the first to fire. For
clarity we plot time only from $t=1$ and do not show the $\sim 1000$
 neurons that were initially ignited
at $t=0$. We
used here $f=0.0033$ and $p_k$ which for $k < \ktail$  is a Gaussian
 centered on $\kbar = 75$ with width $\sigma = 31$ and is a power law
  $p_k \sim k^{-2}$ for $k>\ktail$. We checked that a simple Gaussian
   $p_k$ gives the same qualitative
results.}
\label{f:ex1}
  \end{center}
\end{figure}

We use the simulation for an initial look at the
recruitment process and to identify those neurons that fire first.
We use a Gaussian distribution to describe the experimental
situation as closely as possible, and put the system near
criticality, \ie, with $f$ barely above $\fs$, to observe a large
range of changes in activity. Figure \ref{f:ex1} shows the degree
values $k$ of the neurons as a function of the time step at which
they first fire. It is evident that the neurons that fire first are
either the ones that were ignited externally or those with high $k$.
This is verified in the lower part of Figure \ref{f:ex1}, in which we
plot, for each neuron, its time of ignition as a function of its
in-degree $k$.
It is obvious that the high $k$ neurons totally dominate the initial
stages of the activity.

\begin{figure}[ht]
  \begin{center}
   \includegraphics[scale=0.4]{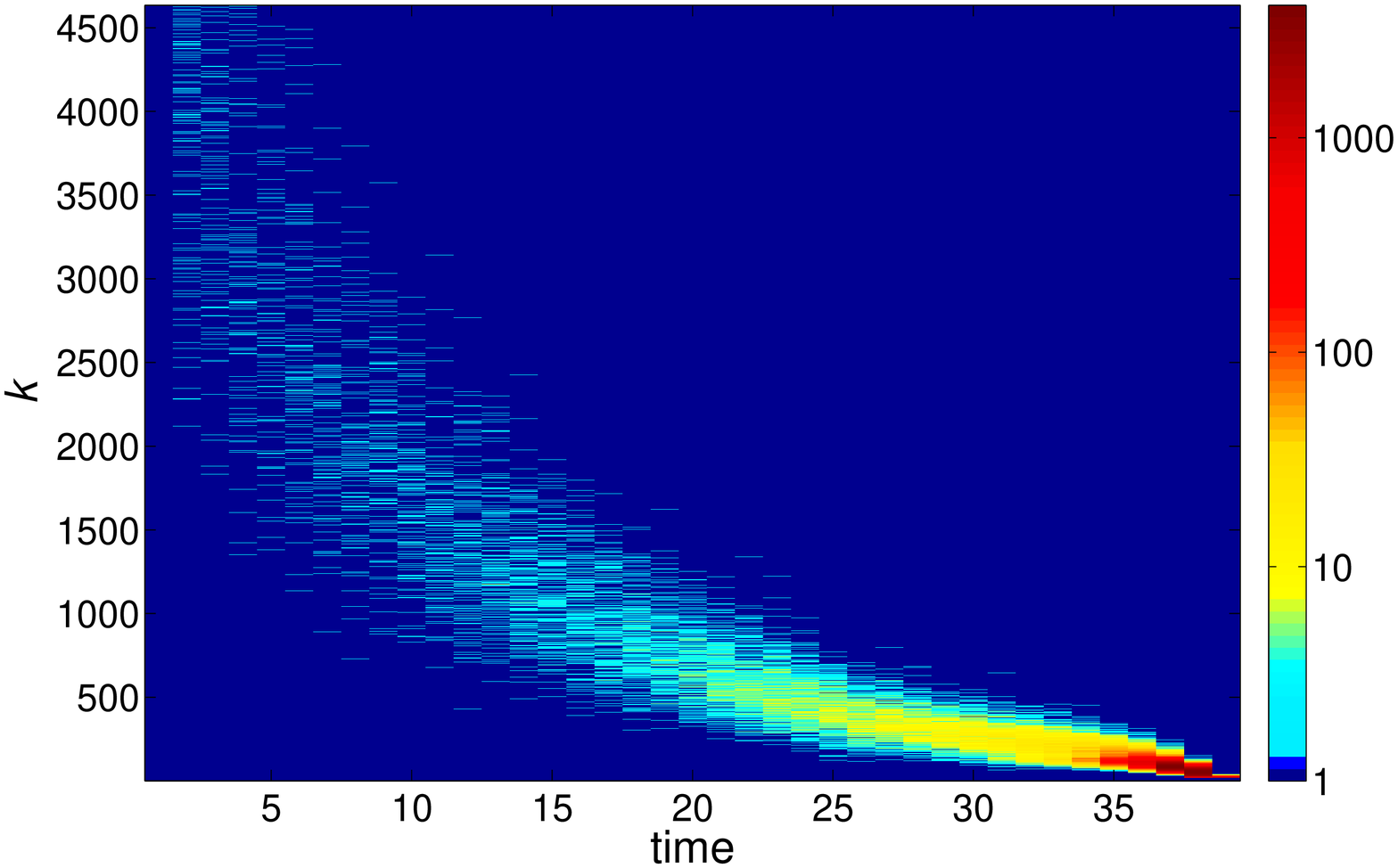}
  \caption{ The logarithmic color coding of the number of neurons with in-degree
  $k$ that ignited at time step $t$.
  The data are the same as in Figure \ref{f:ex1}. }
\label{f:ex1a}
  \end{center}
\end{figure}

Further information on the distribution of ignition times for
different neurons with different in-degree $k$ is given in the
colored map format of Figure \ref{f:ex1a}. The huge majority of
neurons has a low $k$ and ignites very late in the burst. The
first-to-fire neurons, or Leaders, are few and have a wider
distribution of in-degree $k$ at a given time step $t$. The
distribution sharpens as the burst advances in time.

To understand this from the model, we concentrate on the neurons
within a given $k$-class, \ie, the group of neurons with $k$ inputs,
and examine the probability of a neuron within this group to fire,
$\Phi_k$. If the average number of inputs $\kbar$ and the threshold
value $m$ are large numbers, then we can neglect the width of the
binomial distribution and approximate the error function $\Psi_k(m,
\Phi)$ by its limit $\Theta(k-
m/\Phi)$. This simplifies the dynamics Eq.(\ref{kDynamicMaria}) into
\be \Phi_k(t+1) = f+(1-f)\Theta \left(k - \frac{m}{\Phi(t)} \right)
=
\begin{cases}
f &\text{for~}  \Phi(t)<m/k\\
1&\text{for~}  \Phi(t)>m/k
\end{cases}~,
\label{app1} \ee
meaning that under this approximation the whole $k$-class fires once
$\Phi(t)$ exceeds $m/k$. In other words, the $k$-classes are ignited
in steps where in each step the classes whose $k$ is in the range
$m/\Phi(t) < k <  m/\Phi(t-1)$ are ignited. Obviously, the first
nodes to be ignited are those with the high $k$. By summation, one
finds the iterative equation
\be \Phi(t+1) = f+(1-f)\sum_k p_k \Theta\left(k - \frac{m}{\Phi(t)}
\right) =  f+(1-f)\sum_{m/\Phi(t)}^{\infty}  p_k = f+(1-f)
P\left(\frac{m}{\Phi(t)} \right)~, \label{app2} \ee
where $P(k) = \sum_{k}^{\infty} p_k$ is the cumulative distribution.
The dynamics of the approach to the fixed point can be graphically
described as the iterations between the curves $\Phi$ and $f+(1-f)
P(m/\Phi(t))$.

The time continuous version of the iteration equation is
\be \dot\Phi(t) = f - \Phi(t) +(1-f)\Psi(m,\Phi(t)) \simeq f -
\Phi(t) +(1-f)P\left(\frac{m}{\Phi(t)}\right)~, \label{Cont} \ee
which can be integrated, at least numerically, to obtain the
dynamics of the system. Within this approximation, the $k$-class firing is given
 by the step-function (\ref{app1}).
For simple collectivity functions, $\Psi$, and simple degree
 distribution $p_k$ (\ref{Cont}) can be integrated
analytically. In more complicated cases, an iterative scheme, as
detailed in Section III.C below, is needed.

Since these neurons are highly connected, they will statistically
be connected to each other as well. For the experimentally relevant
case of a Gaussian distribution peaked at $78$ connections with a
width of $25$ \cite{PRL,PNAS}, more than $10\%$ of the neurons have over $100$
connections, while about $1\%$ of the neurons have $120$
connections or more.  Since high-$k$ neurons have more inputs and hence a
larger probability to receive inputs from other high-$k$ neurons,
these highly connected neurons form an interconnected subgraph. We
summarize our understanding by stating that \textit{Leaders are
highly interconnected, homogeneously distributed and form a sparse
sub-network}.

In the  Multi Electrode Array experiment about 60 neurons were
monitored, and a burst was observed to begin with one or two of
these neurons. From these initial sites the activity spread.
Identifying these neurons as Leaders, we reach the conclusion that
in every experimentally accessible patch of the network that we
monitor there is a small number of neurons that lead the other
neurons in activation. We therefore deduce from the theory that they
are part of this highly interconnected, sparse sub-network. In the
initial pre-burst period nearby neurons are recruited by inputs from
the Leaders, while in the burst itself all the neurons fire
together. During the pre-burst a spatial correlation to the Leader
exists in its near vicinity, which vanishes as the activity transits
to the burst.

We remark here that within our model a node that fires early is
highly connected. However, the number of connections $k$ and the
threshold for firing $m$ are two factors with the same effect, and
they could in fact interplay to cause a more complex behavior than
we are describing \cite{ZbindenThesis}. One alternative model could
hold the number of connections fixed for all neurons, and only allow
a variation in the number of inputs needed for a neuron to fire $m$.
This would clearly bring about a variety of response times of
neurons, and could create a subgroup of nodes that fire early. If we
allow a few neurons to have a low threshold $m$ then those neurons
will qualify as our Leaders. While there is no evidence to point to
a wide variability in the threshold of the neurons, there are clear
arguments why some neurons may change their threshold in response to
the activity of other neurons, either reducing the sensitivity
(adaptation) or increasing it (facilitation). In Section
\ref{Marumiya} we discuss such a possibility, giving a demonstration
of how such a scenario could evolve.

\subsection{Deducing the connection distribution from the initial growth rates}
\subsubsection{The experimentally relevant case of an exponential
pre-burst}

If the firing order of the neurons is determined by their
connectivity, then by observing and analyzing the evolution of the
burst we may learn about the connectivity of the neurons. We focus
on the experimental fact that the growth rate of the very first
firing is exponential, which leads us in the next sections to
analyze a particular form of the degree distribution that can lead to
exponential growth dynamics.

Our observation that the initial growth of the burst is totally
dependent on nodes at the very high-$k$ side of the degree
distribution gives an opportunity to find the origin of the initial
exponential growth $A(t)=e^{\alpha t}$ observed by Eytan and Marom
\cite{EandMJN}. This regime appears at the very beginning of the
burst (\ie, at the pre-burst defined in \cite{NJP}), and ends when
the majority of the network begins to be active and the actual burst
(also defined in \cite{NJP}) occurs. During this period the
amplitude of activity $A(t)$ grows by a factor of about $30$, and
the value of $\alpha$ is about $0.04-0.05$ kHz ($\alpha$ depends on the
time step chosen, which is taken to be a millisecond in the
experiment \cite{EandMJN} ).  After the exponential regime comes a phase of faster
growth rate, during which the amplitude increases by another factor
of about $10$. The errors on these factors, taken from Eytan and
Marom ~\cite{EandMJN}, are estimated to be no more than $10\%$. We
note that the same exponential growth rate is observed in the
experimental data of Jacobi and Moses~\cite{ShimshonJNP}.

If $\Phi(t)$ is known then one can, in principle, extract the in-degree
distribution $p_k$, since in the random graph scenario nodes with
higher $k$ ignite the next lower level, of $k-1$-nodes. In
particular, as we shall now show, an exponential growth rate is
obtained for the power law distribution $p_k = B k^{-2}$. We begin
by plugging into the approximate dynamics (\ref{Cont}) an
exponential time dependence $\Phi(t)=fe^{\alpha t}$,  where $f$ is
the initially lit fraction, and get:
\be  \dot{\Phi} = \alpha \Phi =
 f - \Phi +(1-f)P\left(\frac{m}{\Phi(t)}\right)~,
 \ee
and therefore
\be
P\left(\frac{m}{\Phi(t)}\right) = \frac{(1+\alpha)\cdot\Phi(t) -f}{1-f}~.
\label{rel}
\ee
Introducing $q=\frac{m}{\Phi(t)}$,
\be
 P(q)=\frac{1+\alpha}{1-f}\cdot\left(\frac{m}{q}\right)-\frac{f}{1-f}~,
 \ee
and we end up with
\be
p_k = -\left .\frac{d}{dq} P(q)\right |_{q=k}
= m\cdot\frac{1+\alpha}{1-f}\cdot\frac{1}{ k^{2}}~.
\ee

\ \

This sets the value for $B = m\cdot\frac{1+\alpha}{1-f}$ in terms of
the growth rate $\alpha$. We find empirically below that the data
are best reproduced for $\alpha=0.04$, impressively close to the
experimental value $\alpha=0.045$. This indicates that the time
steps used in the simulation and in the experiment are similar \ie,
the firing time of a neuron (simulation time) is very close to a
millisecond (experimental time).

\subsubsection{The full degree distribution $p_k$}

However, the distribution obtained above would give an exponential
growth at all times until the whole network is ignited and
$\Phi(t)=1$. That is not the experimental situation. In the data of
Eytan and Marom and in that of Jacobi and Moses \cite{EandMJN,NJP}
the exponential regime includes a small fraction of the nodes (about
$10\%$). It is followed by a faster growth rate, during which the
remaining nodes fire. We furthermore have measured with the
percolation experiments \cite{PRL,PNAS} that the distribution in a
typical culture is well described by a Gaussian, centered on
$k_{\rm{cen}}\simeq 78$ and with a width of about $\sigma = 25$. The
average connectivity, as measured by the mean of the Gaussian, was
shown to increase with the density of plating of the neurons
\cite{PNAS}. We note that these percolation experiments measure the
fixed point of the firing dynamics and are therefore insensitive to
any fat tail of the degree distribution, which governs the
pre-burst.

This leads us to the following tailored solution, which combines
both these experimental inputs and solves the growth rate problem.
We keep the Gaussian distribution for $p_k$ over a large proportion
of the nodes. We have some intuition on why the input degree
distribution of nodes should be Gaussian: it is essentially
determined by the area of the dendritic tree times the density of
the axons that cross that area  (see Discussion). Both the area and the density are
expected to be random variables in a culture grown on a dish. These
random values form a Gaussian distribution with a mean and variance
that are set by biological processes.

For simplicity and conformity with the experimental situation, we
also demand that no node has in-degree less than a minimal $\kmin >
m$, where $m$ is the number of inputs that need to fire for a node
to be excited. We therefore begin with the following distribution
for small k ($\kmin< k < \ktail$): \be p_k \sim \exp \left( -
\frac{(k-k_{\rm{cen}})^2}{2\sigma^2} \right)~. \ee

At high $k$ we need to change to a power law distribution $p_k=Bk^{-2}$,
and we do this from a degree $\ktail$. The value of $\ktail$, among
other parameters, is determined by external considerations along with
consistency constraints, as detailed below.

\subsubsection{Quantitative comparison of model and experiment
 - setting the parameters}
An impressive experimental fact is the large dynamic range observed
in the amplitude of the burst, about two and a half decades in
total. In the experiment, the amplitude grows in the exponential
pre-burst phase by a factor of about $30$, and in the burst
itself by a further factor of about $10$.

In the experiment the large dynamic range and high precision are
obtained by averaging over a large number
of bursts, and can be reproduced in the simulation only if there is
a very large range of available $k$ values, or else the cascade
during which successive $k$-nodes ignite each other does not last
for long enough. To be concrete, we find that we need $\Phi$ to
start from about $f=0.003$ in order to see the amplitude increase by
a factor of $300$ in total. For the growth to be extended in time
and to allow sufficient resolution in the simulation, we demand also to
be near criticality, \ie, $f\simeq \fs$. This slows the process by
adding only a few neurons that ignite at every time step. To obtain
such a very long growth time at any other point, away from
criticality, would require a larger range of $k$, so by staying near
criticality we are actually limiting the range of $k$ to the minimum
necessary to reconstruct the experimentally determined dynamics.

 To ensure that during the  exponential
regime $\Phi$ increases by a factor of $30$ while during the faster growth it grows by
a factor of $10$, the transition from exponential growth to the faster, full blown
firing of the network  is designed to occur at $\Phi=0.1$ and $f$ is set
at $0.0033$.

Since the entry of the $k$ degree node occurs at a
$\Phi=\Phi_k\simeq m/k$, at the transition from pre-burst to
burst we have $\ktail \simeq m/\Phi$. Inserting $\Phi=0.1$ and
$m=15$ gives a characteristic value of $\ktail \simeq 150$. This is
a considerable distance from the peak of the Gaussian, so that it is
justified to describe the majority of the nodes by the Gaussian
distribution.

\begin{figure}[ht]
  \begin{center}
   \includegraphics[scale=0.7]{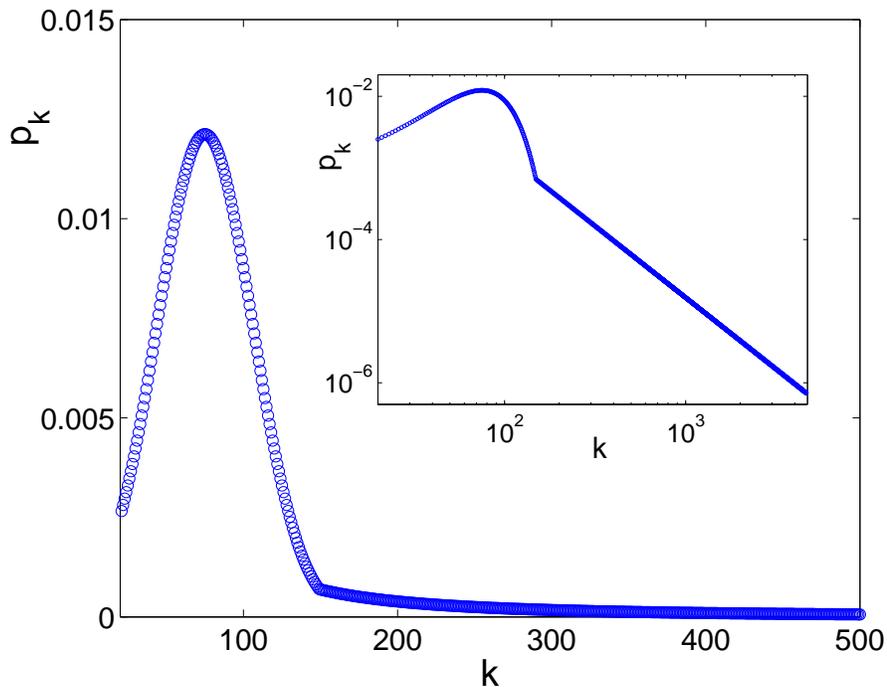}
  \caption{ The in-degree distribution, $p_k$, plotted in both linear (main
plot) and in log-log (inset) coordinates.
Parameters used for the Gaussian are:
  $k_{\rm{cen}}=75$, $\sigma=31$, $k_{\rm{min}}=20$ while the power law tail
  $p\sim B\cdot k^{-2}$ goes from $k_{\rm{tail}}=150$ to $k_{\rm{max}}=4,680$
  and its pre-factor is
  $B=15.65$. This normalizes the distribution to integral 1.
 The  log-log scale in the inset highlights the power-law tail, while the linear scale
of the main plot accentuates the Gaussian that dominates the majority
  of the population.}
\label{f:ex2}
  \end{center}
\end{figure}

The highest cutoff of the degree distribution is in turn determined
by the constraint on the integral over the distribution from
$k_{\rm{tail}}$ to $k_{\rm{max}}$, which should yield a total
fraction of $0.1$, since that is the part of the network that will
ignite in the initial, exponential regime,
$P(\ktail)=\Phitail=0.1$.
This condition allows us to normalize the cumulative function
\be
P(k) = \Phitail \cdot\frac{k^{-1}-\kmax^{-1}}{\ktail^{-1}-\kmax^{-1}}~.
\ee
Plugging this $P(k)$ into (\ref{rel}) yields
\be
P\left(\frac{m}{\Phi} \right)= \frac{\Phi(1+\alpha)-f}{1-f}=
\Phitail \cdot\frac{\Phi/m-\kmax^{-1}}{\Phitail/m-\kmax^{-1}}~,
\label{rel2}
\ee
where we used $\ktail = m/\Phitail$. At $k=\ktail$ Eq.(\ref{rel2})
yields
\be \Phitail=\frac{f}{\alpha+f}~. \ee
Since we have seen that $\Phitail=0.1$, this sets consistency
demands on $\alpha$ and on $f$. For $\Phi(0)=f$, we find after some
algebra:
\be \kmax = m \cdot \frac{1+\alpha}{f}~. \ee
Since the experimentally relevant values of both $f$ and of $\alpha$
(measured in the appropriate time-step)
are known and obey $f \ll \alpha \ll 1$, the approximate relations
are $\Phitail \simeq f/\alpha$ and $\kmax \simeq m/f$.
 This then sets the pre-factor of the distribution $B
= m \cdot \frac{1+\alpha}{1-f}\simeq m$. We end up with the
probability distribution function shown in Figure \ref{f:ex2}, in
which a power law tail from $k_{\rm{tail}}$ all the way to
$k_{\rm{max}}$ is glued onto a Gaussian curve centered on
$k_{\rm{cen}}= 75$.

\begin{figure}[ht]
  \begin{center}
   \includegraphics[scale=0.5]{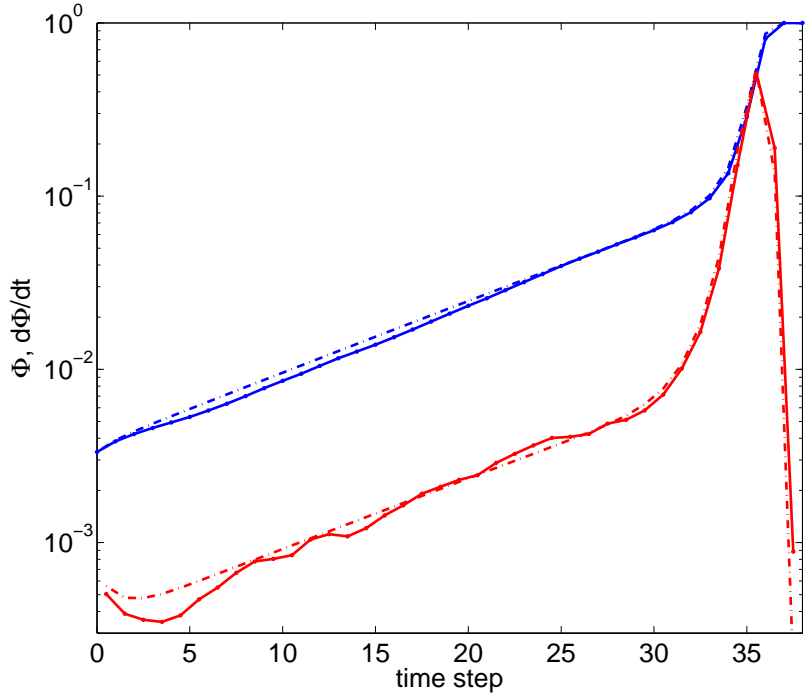}
  \caption{Growth rate of burst for tailored distribution from both
    numerical
simulation using $500,000$ neurons (solid lines) and
iteration of Eq.(\ref{DynamicMaria}) (dashed lines).
 Blue lines show overall firing fraction $\Phi$ for each time step.
 The red curves show the numerical derivative, $\dot{\Phi}$.
The initial slowing down (the dip in the derivative) is due to a
clearly evident ``bottleneck'' in the simulation, during which the
firing almost ceases to propagate.
  The parameters used are $f=0.0033$, $\alpha = 0.03$, and $\kmin =
\rm{round}\left(m(1+\alpha)\right) = 16$, $\kmax = \rm{round}\left(
m(1+\alpha)/f \right) = 4680$. }
\label{f:ex4a}
  \end{center}
\end{figure}

Figure \ref{f:ex4a} shows our main result, in which an exponential
growth rate is reproduced in a simulation employing the in-degree
distribution of Figure \ref{f:ex2}. This exponential phase is
followed by a super-exponential phase, in which the majority of the
network ignites. The majority has in-degree defined by the Gaussian
distribution and therefore they fire practically simultaneously.
Since we tailored $\Phi(t)$, the growth during the exponential phase
is indeed by a factor of $30$, similar to the experimental one.
However, the experimental graphs describe the momentary activity
while $\Phi$ is the total fraction of active neurons and these do
not turn off. The experiment is therefore probably better described
by the derivative of $\Phi$, shown in red. It can be clearly seen that in
fact $\Phi$ and $\dot{\Phi}$ behave very similarly.

We can also compare the simulations of the network with the
numerical solution of the model. For this we use the iterative
scheme defined by Eq.(\ref{DynamicMaria}) to propagate the activity of
the network.
This is shown (in dashed lines) in Figure \ref{f:ex4a}.
The excellent congruence of the simulation and the mean-field equation gives
verification for the use of our  mean field model. The success relies on the absence of large deviations and insignificance of fluctuations, which is true in our model and experiment, due to the benign behavior of the degree distribution and the large number of participating neurons.  The only deviation from this agreement is at the initial steps, where only a small number of leaders are firing.

In summary, from the quantitative comparison we find that the model
has an exponential initial transient if the in-degree distribution
is mostly Gaussian, with $10\%$ of the neurons in the power law
tail, and that the highest $k$ can be in the thousands.

When comparing these results with the experiment, we should remember
that only 60 electrodes are being monitored. The exponential behavior
that is observed over a large dynamic range, can be resolved since
multiple
firings at the same electrode are observed with a resolution better
than 1ms. In the simulation, in contrast, this is modeled by going to
high numbers of neurons, each of which can only fire once.

\subsection{Excitation-dependent threshold}\label{Marumiya}

At the end of Section \ref{IIIA} we noted that the sensitivity of
neurons can be changed\ either by varying the number of their
inputs, or by varying their threshold. Up to now, we have assumed
that the firing threshold in the neural network $m$ is a constant
that does not change as the burst develops, and varied the degree
distribution instead. In this section, we examine the impact of
keeping the connectivity distribution static, while ``loading'' the
recruitment dynamics onto $m$ by making it a dynamical variable that
depends on the history of neuronal activation. Since varying either
parameter ($m$ or $p_{k}$) will lead to the same results, in
principle one could then have any distribution $p_k$ of input
degree, and compensate by varying $m$. One would then have to verify
that the necessary variations in $m$  are biologically reasonable
and feasible. In the experiment this happens via the competing
processes of adaptation and facilitation. Since adaptation would
work opposite the trend observed in the experiment, we discuss only
the possibility of facilitation.

Facilitation of activity can occur if neurons that are already
excited several times are easier to excite at the next time. By
synaptic facilitation we mean the property of a synapse increasing
its transmission efficacy as a result of a series of high frequency
spikes.  We examine here some of these effects by introducing, for
the sake of simplicity, a threshold which is a function of the
average firing state, $m(\Phi)$.

Given the time series of the firing fraction $\Phi(t)$ and a
presumed degree distribution $p_k$, one can invert equations
(\ref{app2}) or (\ref{Cont}) to obtain
\be m(\Phi(t)) = \Phi(t)\cdot K \left( \frac{\Phi(t) +\
\dot\Phi(t) - f}{1-f} \right)~, \label{mphi} \ee
where $K(P)$ is the inverse to the cumulative function
$P(k)=\sum_{k'=k}^{\kmax}p_{k'}$ (such an inverse function exists since
$P$ is monotonic).

It is particularly interesting to ask if the power law degree
distribution $p_k \sim k^{-2}$, supports a biologically feasible
form of $m(\Phi)$, following the initial exponential regime where
$m$ should be constant. In this case the cumulative function is
\be
P(k)=\frac{k^{-1}-\kmax^{-1}}{\kmin^{-1}-\kmax^{-1}}~,
\ee
where
$\kmin$ and $\kmax$ are the limits of the distribution. The inverse
function is \be
K(P)=\frac{1}{P\cdot(\kmin^{-1}+\kmax^{-1})+\kmax^{-1}}~. \label{KP}
\ee

To further advance we need to model the burst itself, which grows
exponentially at first, then grows even faster, at a super exponential
rate and finally saturates when all the network has fired. This
behavior can be described by the function
\be \Phi(t) =f\cdot\ \frac{1-e^{-\alpha\ts}}{e^{-\alpha t} - e^{-\alpha\ts}}~, \ee
with $\ts$ a parameter to be determined from comparison to the
experiment. This kind of burst function starts as an exponential
$\Phi(t) \sim e^{\alpha t}$ and begins to diverge after  $\ts $ time steps,
but reaches $\Phi(t) = 1$
slightly before fully diverging, at $t_1 = \ts - \alpha^{-1 }
\log[1+ f(e^{\alpha \ts} - 1)]\simeq \ts - (f/\alpha)\cdot e^{\alpha\ts}$.

For this profile
$\dot{\Phi} = \alpha \Phi[1+\Phi/(f(e^{\alpha\ts}-1))]$. Plugging into
(\ref{mphi}) yields
 \be\label{mphi1}
m(\Phi(t))=\frac{\kmin} {1+\alpha
\left[1+\Phi(t)/(f(e^{\alpha\ts}-1))\right]}~. \ee

\begin{figure}[ht]
  \begin{center}
   \includegraphics[scale=0.5]{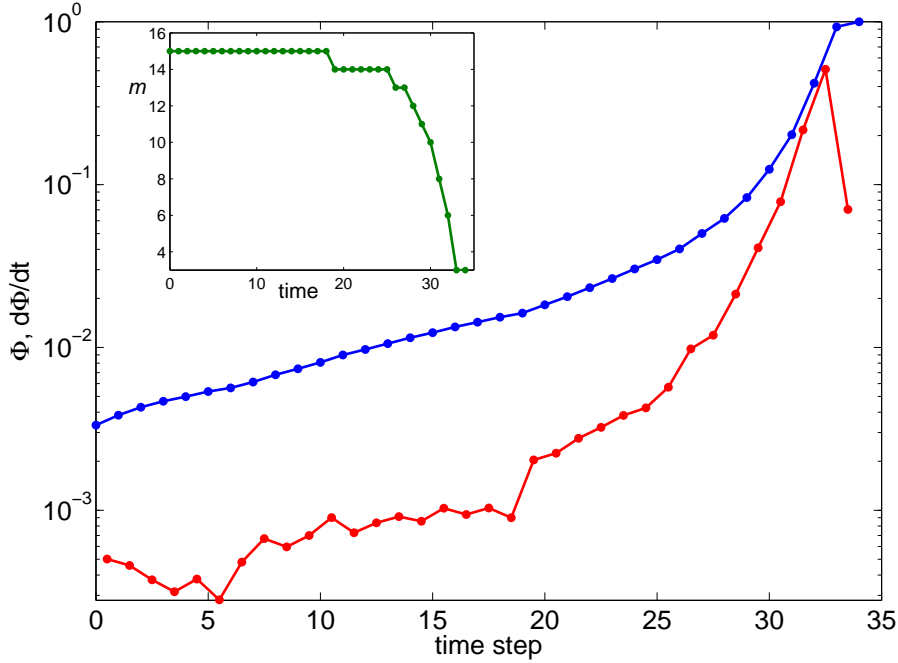}
\caption {Growth rate of burst for $k^{-2}$ power
law distribution and variable $m(\Phi(t))$ using Eq.(\ref{mphi}). The curve
is calculated from numerical
   solution of the iteration Eq.(\ref{DynamicMaria}).
 Blue line shows overall firing fraction $\Phi$ for each time step.
 The red curve shows the numerical derivative, $\dot{\Phi}$. The
 parameter $\ts$ used is $\ts = 40$ time steps, and all other
 parameters are as in Figs. \ref{f:ex2} and \ref{f:ex4a}.
 Inset: The threshold $m(\Phi(t))$ decreases during the simulation as $\Phi$
increases according to (\ref{mphi1}). The value of $m$ used in the
simulation is the integer part of (\ref{mphi1}) hence the discrete
jumps in its value. Until about $t=18$ the value $m$ is unchanged
and the $\Phi(t)$ and $\dot{\Phi}(t)$ profiles are exponential. Then
$m$ starts to decrease sharply and induces super-exponential growth
of $\Phi(t)$.} \label{maromia}
  \end{center}
\end{figure}

We see that $m(\Phi)$ starts from $m(f)\simeq \kmin/(1+\alpha)$ and
ends at $m(1) \simeq \kmin /[1+\alpha/(f(e^{\alpha\ts}-1))]$. Therefore, $m$
decreases by a ratio of $m(1)/m(f) \simeq
[1+\alpha/(f(e^{\alpha\ts}-1))]^{-1}$,
which for the experimental parameters is around $5$, \ie, from $m(f)
\simeq 15$ to $m(1)\simeq 3$, a biologically
reasonable variation. The actual value of $m$ that we use in the
simulation is that of the nearest integer obtained by rounding
Eq.(\ref{mphi}).
Figure \ref{maromia} shows the behavior of the burst as
a function of time for the power law distribution $p_k \sim k^{-2}$
with variable  $m(\Phi)$.


\subsection{Summary of Results}

In summary, we have shown here that the experimental situation of an
exponential transient followed by super-exponential growth can be well
described in our model of an in-degree distribution that is $k^{-2}$ at
high $k$ but is Gaussian for the majority of the neurons. The
initial transient of an exponential is determined directly by the
power law tail of the degree distribution of Leaders.

The $k$-degree values needed to describe the data reach a maximum value that
is many tens of standard deviations from the mean.
Although the average value of the degree distribution remains in the region of
$100$, a few neurons (in a network of a million nodes) can have
thousands of connections.

We remain with the question of how significant is the need for an exponent
${-2}$ in the power law distribution, and whether small deviations will
change the exponential growth rate. Is there any logical or biological reason for this
power law to be built up?

On the experimental side, the search for a few highly connected
neurons would be needed. One possibility is that Leaders are neurons
of a different species then that of the majority. Identifying them,
investigating their properties and potentially intervening by
disrupting their function are all important experimental goals.

\section{Discussion}

\subsection{First-to-fire neurons and Leaders}

In \cite{NJP, ZbindenThesis, ZbindenlIF} Leaders were defined
through an intricate mathematical procedure. In particular, this
definition allowed for exactly one Leader per burst, which ignites a
pre-burst, and then a burst. In the present paper, a simpler
definition is used, which amounts to take into account basically all
neurons which fire at the beginning of activity right after the
initial fraction $f$. Since in the initial period of the burst there
are only very few neurons active, the development of the burst
depends critically on those neurons.

Within the QP model, high $k$ neurons activate the low $k$ neurons.
So the highest $k$ neurons are the ones that trigger the burst.
It follows that some of the highest $k$ (in-degree) neurons are both
first to fire and Leaders.

Looking only at in-degree is only part of picture. Indeed, high
$k$-classes are ignited first. However, their contribution to the
firing propagation depends on the out-degree. Nodes with no outputs
may fire early but contribute nothing to the ignition of others.
Nevertheless, since we assumed that the in- and out-degree are
uncorrelated, Leaders are among the early igniting nodes.

\subsection{How can we get a distribution of in-degree which is Gaussian with a
$k^{-2}$ tail?}

An interesting question is what kind of growth and development
process would lead to a distribution of in-degree that is
essentially a Gaussian centered on a value of about $k_{\rm cen}=75$,
but has a tail that goes like $\kin \sim k^{-2}$, and can reach
in-degree in the thousands, $\kmax\sim 3,500$.

We propose the following simplified and intuitive geometrical
picture for how $\kin $ and $\kout$ are determined. Each neuron in
the culture has a spatial extent that is accessed by its dendrites
(the ``dendritic tree'') and characterized by a length scale $\ell$.
The dendrites have no a-priori preferred direction, and the
dendritic tree is typically isotropic and characterized by a length
scale $r$. Axons, on the other hand, go off in one direction, and
their length determines the number of output connections the neuron
will have. The dendritic tree is ``presented'' to axons of other
neurons. If the axon of a neuron happens to cross the dendritic tree
of another neuron then, with some fixed probability (which we take
for simplicity to be unity), a connection is made between the two
neurons. The number of {\bf in}-connections is therefore related to
the size of the dendritic tree and to the number of axons crossing
it, \ie, the density of axons. The number of {\bf out}-connections
of a neuron is determined by the length of its axon, the size of the
dendritic tree of other neurons and the density of neurons.

\begin{figure}[ht]
  \begin{center}
   \includegraphics[scale=0.55]{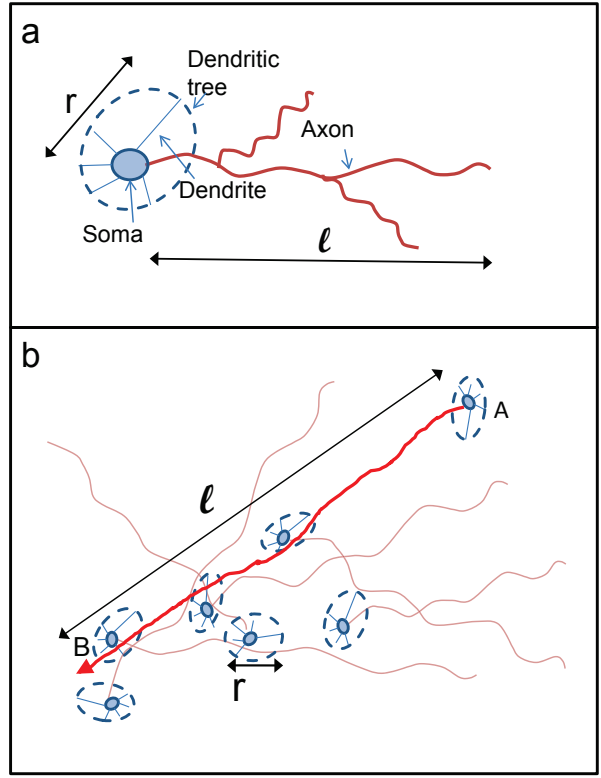}
\caption {Schematic picture of the relation between axon and
dendrite lengths $\ell$ and $r$ to the number of connections $\kin$
and $\kout$.\ (a) Two lengths characterize the connections of each
neuron: its axonal length $\ell$ and the typical size of its
dendritic tree $r$. While the dendritic tree is in general expected
to be homogenous, it can have dendrites that go off much farther
than the others. (b) A connection from Neuron A to Neuron B will be
made if the trajectory of the axon extending from Neuron A will
intersect the dendritic tree of Neuron B. The probability for that
to happen depends on the probability $P(\ell)$ for A to have an axon
of length longer than $\ell$ and on the probability $p(r)$ for B to
present a dendritic tree of cross section $r$.} \label{Lolipop}
  \end{center}
\end{figure}

There are two corresponding length distributions $p(\ell)$ and
$p(r)$ and a density $n$ that determine the number of connections.
$p(\ell)$ and $p(r)$ are the probability distribution of the axon
and dendrite lengths respectively, while $n$ is the density of
neurons per unit area.

The number of in-connections of a neuron is obtained by calculating
the probability of an axon emitted from another neuron located a
distance $\ell$ away to cross its dendritic tree. To get the number
of connections $\kin $ for a neuron with dendritic tree of size $r$
we look for the axons that will cross one of its dendrites: \be \kin
(r) = n \int_0^\infty d\ell  2\pi \ell \frac{2r}{\ell} P(\ell)= 4\pi
n r \int_0^\infty d\ell P(\ell)~. \ee

Here $P(\ell)=\int_\ell^\infty p(\ell')d\ell'$ is the cumulative sum
of probability that the length of an axon exceeds $\ell$ (since it
would then cross the dendritic tree). We ignore the slight $r$
dependence of the lower limit of the integral. $n$ is the density of
neurons per unit area (about $500$ neurons per mm$^2$), and
$\frac{2r}{\ell}$ the angle extended by the dendritic tree as seen
from the axon's neuron of origin.

We now insert for $p(\ell)$ the Gaussian with power law tail:

$p(\ell)= A \cdot e^{-\frac{(\ell-\ell_0)^2}{2\sigma^2}}$ if $\ell <
\ltail$ and $p(\ell)= B \cdot \ell^{-2}$ otherwise, with $A\gg B$
normalization factors.

For $\ell < \ltail$ we get \be P(\ell)= A \int_{\ell}^{\ltail}
e^{-\frac{(l'-l_0)^2}{2\sigma^2}} dl'+ B\int_{\ltail}^{l_{\rm{max}}}
(l')^{-2} dl' = \const - A\cdot \erf(\ell)~, \ee while for
$\ltail<l<\lmax$: \be P(\ell)= B\int_{\ell}^{\lmax} \ell'^{-2}
d\ell' = B\left (\frac{1}{\lmax}-\frac{1}{\ell}\right)~. \ee The
integral over $P(\ell)$ gives one constant term and one that goes as
$\log(\lmax)$. Since the maximal length is determined by the size
$L$ of the culture dish, we remain with a term of $\log(L)$.

We  get that the number of in-connections is
 \be
\kin  (r)\simeq  n \cdot r \cdot \log(L)~. \ee

We are now in a position to ask where the tail of high connections
arises. In principle, it could arise from fluctuations in the
density $n$. The neural density is theoretically determined by
throwing down a random set of points on the plane. In the experiment, the
neuronal cultures do not exhibit  clustering to such a degree that would
induce so strong a fluctuation with such high density. To get the necessary
range of a factor of $10-30$ in $k$ values that the theoretical
explanation of the experimental data point to, we would need the
density to change in a similar manner. This seems unrealistic. A
further strong argument against fluctuations in density is that
these would lead, in our picture, to a high number of input
connections in one single neighborhood. In particular, this would
lead to many more high-$k$ neurons than the distribution allows for.

Thus we are led to the conclusion that the power law tail of
$\kin $ has its origin in the distribution of dendritic trees
$p(r)$. While the typical dendritic tree is probably a circle of
with radius $r=100-200$ micrometer, it may have outgrowths in one or
more directions that reach as far as $l=1-2$ mm but can, with small
probability, go as far as the dish size $L$.

\subsection{Trading off static connectivity distribution for dynamical threshold}

While the possibility of exchanging between elementary dynamics and
connectivity statics does not come as a surprise, the lesson we take
from the results of Section \ref{Marumiya} is important indeed: One
cannot distinguish, by observing network dynamics in and by itself,
between a static connectivity-based mechanism and a mechanism that
employs dynamics at the elementary level. While we tend to believe
it is the connectivity that is dominant, one cannot rule out the
neuron's internal processes as a possible explanation for the
dynamics.

When only a small fraction $\Phi$ of the network is active, the
chances that a given neuron is activated at a high frequency are
low. Hence, chances of changes in threshold as a result of synaptic
or membrane dynamics are low. However, as the active fraction $\Phi$
of the network grows, the chances of a neuron to be bombarded at
high frequency become higher; we could then get a dependence
$m(\Phi)$, since changes in threshold (i.e. membrane dynamics) or
synaptic efficacy (e.g. facilitation) are expected.

In many studies of biological networks, this ambiguity is somewhat
neglected in favor of a more static view, largely due to lack of
access to elementary level dynamics. In the case of neuronal
excitability, single element dynamics is experimentally accessible,
and the existence of dynamical-thresholds are well documented. Our
results in Section \ref{Marumiya} indicate that this is a sufficient
explanation to the phenomenon of an early exponential recruitment
rate followed by faster growth process.

\subsection{The effect of limiting size: One-dimensional cultures}
 Initiation of activity in one-dimensional cultures  seems to be very different
from  the Leaders scenario.  In one-dimensional cultures, we
\cite{OferJNP2} have shown that the activity originates locally, at
well defined ``Burst Initiation Zones'' (BIZs) that have a limited
spatial extent. There are usually a small number of such BIZs,
typically one or two per centimeter, that operate independently of
each other. The BIZs are characterized by a high density of
excitatory neurons and a low density of inhibitory ones. Firing
activity that originates in a BIZ will propagate out as a wave-like
front with a constant velocity, and invade the rest of the culture
until all neurons have fired.

One explanation for the difference in behavior of BIZ and Leaders is
in the dimensionality. However, the basic argument we presented for
the number of input connections relates it to the multiplication of
the area of the dendritic tree by the density of the axons that
cross through this area. Since both the radius of the dendritic tree
and the width of the line are on the order of 100 micrometer, there
should be no difference in the first factor. As for the density of
axons, there is no direct information, but also no compelling
argument why 1D cultures should differ in this from 2D cultures.

A different possibility, and the one we believe to be correct, is
just that there are too few neurons in the culture \cite{Remark}. That would
impact on any small culture, both 2D and 1D. Changing the number of
neurons has the largest effect on the realization of the tail of the
probability distribution $p_k$, since high $k$ values have a low
probability and will not be obtained. This can completely disrupt
the form of the degree distribution. In turn, it also affects the
value of $\fs$, the fraction of initial firing needed for ignition
of the full culture.

\begin{table}
  \begin{tabular}{|c|c|c|c|}\hline
$\fs$ & N & $k_{\rm{max}}$ theory & $k_{\rm{max}}$ realized \\\hline
0.05 & 500     & 312 & 237 \\\hline 0.022  & 5,000   & 709 &
615\\\hline 0.0076  & 50,000  & 2,053 & 1,795
\\\hline 0.0051 & 100,000 &2,836 & 2,512 \\\hline
0.0038 & 500,000 & 4,680 & 4,680\\\hline
\end{tabular}
\caption{Study of finite size networks: For small $N$
  networks, $f_*$ needs to be larger and the
  experimentally accessible
  $\kmax$ does not reach the theoretical prediction. This shows that
   the ignition process is less
  efficient than for large $N$.
}\label{t:thetable}
\end{table}

The results of simulating of excitation in varying numbers of
neurons are given in Table \ref{t:thetable}:

We immediately see that indeed $\fs$ depends strongly on $N$. A power law
fit indicates that $\fs \sim N^{-1/2}$. At about
$N=200,000$ the curve flattens out, and reaches the theoretical
($N=\infty$) value.
The reason for this originates in the constraints imposed between $\fs$ and
$k_{\rm{max}}$, $\fs\simeq\frac{m}{k_{\rm{max}}}$.
In any realization of finite size $N$, any $k$ with $p_k < 1/N$ is very unlikely
to  be observed. Since  $p_k \sim k^{-2}$ both $\kmax$ and $\fs$ are constrained
by the $N^{-1/2}$.
We can conclude that within the Quorum Percolation model smaller
cultures require a much larger fraction of initial activity to
sustain a burst.

\acknowledgments{We are indebted to Shimon Marom for many
stimulating discussions and in particular for suggesting the
equivalent effect  of in-degree distributions and neuronal
thresholds. We also thank Shimshon Jacobi, Maria Rodriguez Martinez
and Jordi Soriano. This research was partially supported by the
Fonds National Suisse, the Israel Science Foundation grants number
1320/09 and 1329/08, Minerva Foundation Munich Germany and the
Einstein Minerva Center for Theoretical Physics. OS acknowledges
support from the German Ministry for Education and Science (BMBF)
via the Bernstein Center for Computational Neuroscience (BCCN)
G\"{o}ttingen (Grant No. 01GQ0430) }

\newpage
\bibliographystyle{apsrmp}
\bibliography{PrimitiveLeadersRef}

\end{document}